# Seebeck effect in the conducting $LaAlO_3/SrTiO_3$ interface

I. Pallecchi, M. Codda, E.Galleani d'Agliano, D.Marré
CNR-INFM-LAMIA, c/o Dipartimento di Fisica, Via Dodecaneso 33, 16146 Genova, Italy

A. D. Caviglia, N. Reyren, S. Gariglio, J.-M. Triscone
DPMC, Université de Genève, 24 Quai E.-Ansermet CH-1218 Genève, Switzerland

**Abstract**

The observation of metallic behavior at the interface between insulating oxides has triggered worldwide efforts to shed light on the physics of these systems and clarify some still open issues, among which the dimensional character of the conducting system. In order to address this issue, we measure electrical transport (Seebeck effect, Hall effect and conductivity) in $LaAlO_3/SrTiO_3$ interfaces and, for comparison, in a doped $SrTiO_3$ bulk single crystal. In these experiments, the carrier concentration is tuned, using the field effect in a back gate geometry. The combined analysis of all experimental data at 77 K indicates that the thickness of the conducting layer is ~7 nm and that the Seebeck effect data are well described by a two-dimensional (2D) density of states. We find that the back gate voltage is effective in varying not only the charge density, but also the thickness of the conducting layer, which is found to change by a factor of ~2, using an electric field between -4 and +4MV/m at 77K. No enhancement of the Seebeck effect due to the electronic confinement and no evidence for two-dimensional quantization steps are observed at the interfaces.

# 1. Introduction

Since the seminal work by Ohtomo and Hwang [1], intensive research has been focused on the realization and investigation of conducting interfaces between wide band gap perovskite insulators. One of the most studied of such systems is the $LaAlO_3/SrTiO_3$ interface, characterized by high mobility at low temperature [2], superconductivity [3] and possible magnetic effects [4,5]. This interface has been almost exclusively fabricated by growing epitaxially a few unit cells of $LaAlO_3$ by Pulsed Laser Deposition (PLD) onto $TiO_2$ terminated (001) $SrTiO_3$ single crystalline substrates. The much debated key controversy concerns the origin of the interfacial electronic system. One possible scenario is based on an "intrinsic" mechanism, known as polar catastrophe. Because of the polar discontinuity occurring at the $LaAlO_3/SrTiO_3$ interface [6], an electronic reconstruction occurs in order to prevent the potential build up across the $LaAlO_3$ thickness and half an electron charge $-e/2$ per areal unit cell is transferred to the interface. The competing scenario, based on an "extrinsic" mechanism, attributes the electronic charge at the interface to ionic defects such as oxygen vacancies created during the $LaAlO_3$ growth [7,8] or cationic interdiffusion of La and Sr at the interface [9].

Some experimental observations are in line with the oxygen vacancy scenario. For instance, $SrTiO_3$ is known to be very sensitive to oxygen vacancies and conducting $SrTiO_3$-based interfaces are all electron-doped systems. Indeed, conducting interfaces are only obtained with $LaAlO_3$ grown onto $TiO_2$-terminated $SrTiO_3$ substrates, whereas conductivity and carrier density decrease proportionally to the SrO coverage from 0 to 1 monolayer [2,10]. The Ti mixed valency is thought to be crucial for obtaining conduction at the $(LaO)^+/(TiO_2)^0$ interfaces [11]. Recently, $LaAlO_3/LaVO_3$ hole-type interface conduction has however been observed [12], without completely sorting out the ambiguity, though. On the other hand, other experimental findings are pointing to the intrinsic polar discontinuity scenario. For example, the existence of a critical $LaAlO_3$ thickness of around 4 unit cells below which the interface are insulating [13,14] or the fact that the sheet conductance for samples of different thicknesses above 4 unit cells (up to about 15 unit cells [15]) is found to be essentially constant, can be understood in the polar catastrophe scenario [13]. At this point, the origin of the electron gas is still open and the debate continues. It is likely that several effects are at work, the deposition protocol playing a key role on the electronic properties of the system [16].

Another related open issue is about the thickness of the conducting layer and its dimensionality. It is nowadays accepted that the growth conditions determine the extension of the electron gas in these heterostructures. For samples grown in high oxygen pressure ($P_{O2}>10^{-5}$ mbar) and oxygen annealed, the analysis of the anisotropy of the superconducting properties in magnetic fields points to a thickness $t \approx 10$ nm [17]. Indications of a two-dimensional (2D) Berezinskii-Kosterlitz-Thouless type superconductivity seem to confirm that the conducting layer thickness $t$ is smaller than the coherence length $\xi$ found to be $\approx 70$-100 nm [3]. Also conducting atomic force investigations have confirmed similar quantitative results about the value of $t$ [16,18]. At room temperature, hard x-ray photoelectron spectroscopy [19] experiments suggest an extension of the gas limited to a few unit cells. At low temperatures, the large dielectric constant of $SrTiO_3$ most probably leads to a delocalization of mobile carriers from ionic charges at the interface significantly enhancing the screening length [8]. On the other hand, in samples deposited at low oxygen pressure ($P_{O2}<10^{-5}$ mbar) oxygen vacancies extend more deeply in the STO substrate, as magnetoresistance oscillations at low temperature and high field have indeed indicated [20,2].

Several different kinds of measurements have been carried out to try answering the issues discussed above, such as magnetoresistance oscillations [20], transport [3], spectroscopic techniques [8,9]. Yet, to date, no measurements of the Seebeck coefficient $S$ have been carried out. The Seebeck coefficient contains information on the Fermi surface and carrier density, so that it could clarify the dimensionality and properties of the interfacial electronic system.

Moreover, the exploration of the Seebeck coefficient in systems with reduced dimensionality has technological implications. In 2D systems, the dimensionless figure of merit $Z=S^2\sigma T/\kappa$, where $\sigma$ and $\kappa$ are the electric and thermal conductivities respectively, has been theoretically [21,22,23,24] and experimentally [25] found to be enhanced as compared to its 3D counterpart, suggesting possible applications in cooling systems. The Seebeck coefficient itself may be potentially enhanced in certain 2D systems, allowing improved thermoelectric performances [25,26,27,28]. Two-dimensional SrTiO$_3$-based systems, which are non toxic, cheap and with simple crystal structure compatible with other multifunctional oxides, have been explored as well for this purpose. Indeed, an enhancement not only in |S| but also in its dependence on the carrier density has been measured in SrTiO$_3$/SrTi$_{0.8}$Nb$_{0.2}$O$_3$ superlattices [29,30]. Field effect modulation of S in SrTiO$_3$-based transistors has also been demonstrated [31].

In this work, we present the first measurements of Seebeck effect in LaAlO$_3$/SrTiO$_3$ interfaces, in the diffusive regime, as a function of temperature and carrier density, the latter being modulated by field effect. Combining these data with measurements of electric conductivity and Hall effect as a function of temperature, either with or without field effect, we extract information on the dimensional character and transport properties of this system. The paper is organized as follows: in section 2 we present the framework for modeling of thermopower, for the cases of 3D and 2D degenerate semiconductors; in section 3 we give details about sample preparation and measurement techniques; in section 4 we present experimental results and we carry out quantitative analysis; finally in section 5 we draw our conclusions.

## 2. Model of diffusive Seebeck coefficient

The diffusive contribution to the thermoelectric power is well described by the Cutler-Mott formula appropriate for degenerate semiconductors [32,33]:

$$S = -\frac{\pi^2}{3e}K^2T\frac{\partial \ln(\sigma(E))}{\partial E}\bigg|_{E=E_F} \approx -\frac{\pi^2}{3e}K^2T\left(\frac{\partial \ln(n)}{\partial E}\bigg|_{E=E_F} + \frac{\partial \ln(\tau)}{\partial E}\bigg|_{E=E_F}\right) \quad (1)$$

where $K$ is the Boltzmann constant, $e$ the positive electron charge, $n$ the density of charge carriers, $\tau$ the scattering time and $E_F$ the Fermi energy. The term related to the derivative of the charge carrier velocity, which is usually neglected, has been omitted. The negative sign applies to electron-type charge carriers, relevant in this work. Depending on the dimensional character of the system and thus on the functional form of the density of states, eq. (1) can be expressed in terms of more straightforward parameters.

### 2.a 3D case

In the 3D case, within the quasi-free electron approximation, the functional form of the density of states $N_{3D}(E) \propto E^{1/2}$ yields $\partial \ln(n_{3D})/\partial E\big|_{E=E_F} = 3/2$, so that eq. (1) simplifies to:

$$S_{3D} = -\left(\frac{3}{2}+\alpha\right)\frac{\pi^2}{3}\frac{K}{e}\left(\frac{KT}{E_F}\right) \quad (2)$$

where α describes the functional dependence of the scattering time on the energy $\tau \sim E^\alpha$ and its value depends on the dominant scattering mechanism. In many cases α≈0 is assumed for simplicity and for lack of a solid theoretical back up. However, it has been calculated [34] that *α=-0.5* for scattering with acoustic phonons, while for most other scattering mechanisms *α* is between *0* and *-1*. Also experimental values are found in this range [35].

The degenerate carrier density per unit volume $n_{3D}$ is given by:

$$n_{3D} \approx \frac{8\pi(2m_{eff}E_F)^{3/2}}{3h^3} \quad (3)$$

where $h$ is the Plank constant and $m_{eff}$ is the electron effective mass. Hence, combining equations (2) and (3), the explicit dependence of $S_{3D}$ on the carrier concentration is obtained:

$$S_{3D} = -\left(\frac{3}{2}+\alpha\right)\frac{8\pi^{8/3}K^2}{3^{5/3}h^2e}m_{eff}\frac{T}{n_{3D}^{2/3}} \quad (4)$$

## 2.b 2D case

For a 2D system, the relationship between the Fermi level and the carrier concentration per unit area $n_{2D}$ is expressed by:

$$n_{2D} \approx \int_0^{E_F} \frac{4\pi m_{eff}}{h^2}\left(\sum_\nu \Phi(E-E_\nu)\right) dE = \frac{4\pi m_{eff}}{h^2}\sum_\nu \left((E_F - E_\nu)\Phi(E_F - E_\nu)\right) \quad \text{with } E_\nu = \frac{h^2 \nu^2}{8 m_{eff} t^2} \quad (5)$$

where, $\nu$ is the index of the discrete 2D energy levels and $\Phi(x)$ is the heaviside step function (whose value is equal to zero for negative argument $x<0$ and equal to one for positive argument $x>0$). For the $\nu$-th energy level $E_\nu$, an infinite rectangular quantum well of width $t$ is assumed. The total Seebeck coefficient $S_{2D}$ is a sum of the Seebeck coefficients of all occupied levels $S_\nu$, weighed by the respective conductivities $\sigma_\nu$:

$$S_{2D} = \frac{\sum_\nu \sigma_\nu S_{2D\,\nu}}{\sum_\nu \sigma_\nu} = \frac{\sum_\nu n_{2D\,\nu} S_{2D\,\nu}}{\sum_\nu n_{2D\,\nu}} \quad (6)$$

The last expression of eq. (6) comes out if the mobility is assumed to be the same for all the levels, independently of the carrier density. The occupation of the $\nu$-th energy level is calculated as:

$$n_{2D\,\nu} = \int_0^{E_F} \frac{4\pi m_{eff}}{h^2}\Phi(E-E_\nu) dE = \frac{4\pi m_{eff}}{h^2}(E_F - E_\nu)\Phi(E_F - E_\nu) \quad (7)$$

and for each 2D level, $S_\nu$ is obtained from eq. (1), keeping into account that $N_{2D}$ is constant as a function of the energy and consequently $\partial \ln(n_{2D})/\partial E\big|_{E=E_F} = 1$:

$$S_{2D\,\nu} = -(1+\alpha)\frac{\pi^2}{3}\frac{K}{e}\left(\frac{KT}{E_F - E_\nu}\right) \quad (8)$$

In the above analyses, the thermal smearing of the Fermi distribution function is neglected; its effect is not crucial for degenerate semiconductors at low temperatures, but contributes in averaging out quantization effects when the Fermi temperature $T_F=(E_F-E_{\nu*})/K$, with $\nu*$ index of the highest occupied level at $T=0$, becomes comparable or smaller than the measuring temperature.

## 2.c Comparison between 2D and 3D cases

In figure 1, we present some examples of Seebeck coefficient calculated using either the 3D approach or the 2D one. In all cases, we fix the parameter $\alpha=-0.5$ and the isotropic effective mass $m_{eff}=4m_0$, which are both reasonable values for SrTiO$_3$, as it will be explained in the following section. In the uppermost panel, the behaviour of $S$ as a function of the quantum well width $t$, at fixed sheet carrier concentration $n_{2D}=1.2\ 10^{14}\ cm^{-2}$ is displayed. On the right-hand axis, the corresponding number of occupied energy levels is also shown. It is clear that in both 3D and 2D cases the absolute value of $S$ decreases with decreasing $t$. In particular, apart from quantization steps, $S_{2D}$ has an average $\propto t^{2/3}$ behaviour, as indicated by the dashed fitting line. This is just the same dependence as $S_{3D}$; indeed, from eq. (4) we get $S_{3D} \propto n_{3D}^{-2/3} = (n_{2D}/t)^{-2/3} \propto t^{2/3}$. In the middle panel, the dependence of the Seebeck coefficient on the sheet carrier density $n_{2D}$ at fixed width $t=7nm$ is presented. Similarly as in the previous case, the 2D and 3D approaches give the same average dependence $\propto n_{2D}^{-2/3} = (n_{3D} \cdot t)^{-2/3} \propto n_{3D}^{-2/3}$. Finally, the bottom panel shows how S changes if $t$ and $n_{2D}$ are varied in such a way that the volume carrier concentration $n_{3D}$ is kept constant. Obviously, according to eq. (4) the 3D case gives a constant $S_{3D}$, but remarkably also in the 2D case we can say that $S_{2D}$ varies in quantized steps around a constant value. In the above described examples, the effect of temperature is omitted; actually, thermal smearing of the Fermi distribution function blurs quantization steps with increasing temperature. In the inset of the middle panel, $S_{2D}$ curves calculated, as in ref. [21,25,36], keeping into account this effect at T=77K and T=10K are also shown. Clearly, at 77K the smearing of steps is severe; nonetheless, when only a few levels are occupied, the 2D approach is more suitable than the 3D one.

The above described behaviors of $S_{3D}$ and $S_{2D}$ are worth some considerations. Noticeably, the same functional dependence of $S_{2D}$ and $S_{3D}$ shows that, even in the 2D case, the thermopower is determined by the volume carrier concentration rather than by the sheet carrier concentration, provided that quantization steps are averaged out. For this reason, the combination of $S(n_{3D})$ and Hall resistance $R_{Hall}(n_{2D})$ measurements allows a direct comparison of $n_{2D}$ and $n_{3D}$ and thus is a powerful tool to get information on the extension of the carrier distribution of a system, which is a relevant issue for LaAlO$_3$/SrTiO$_3$ interfaces.

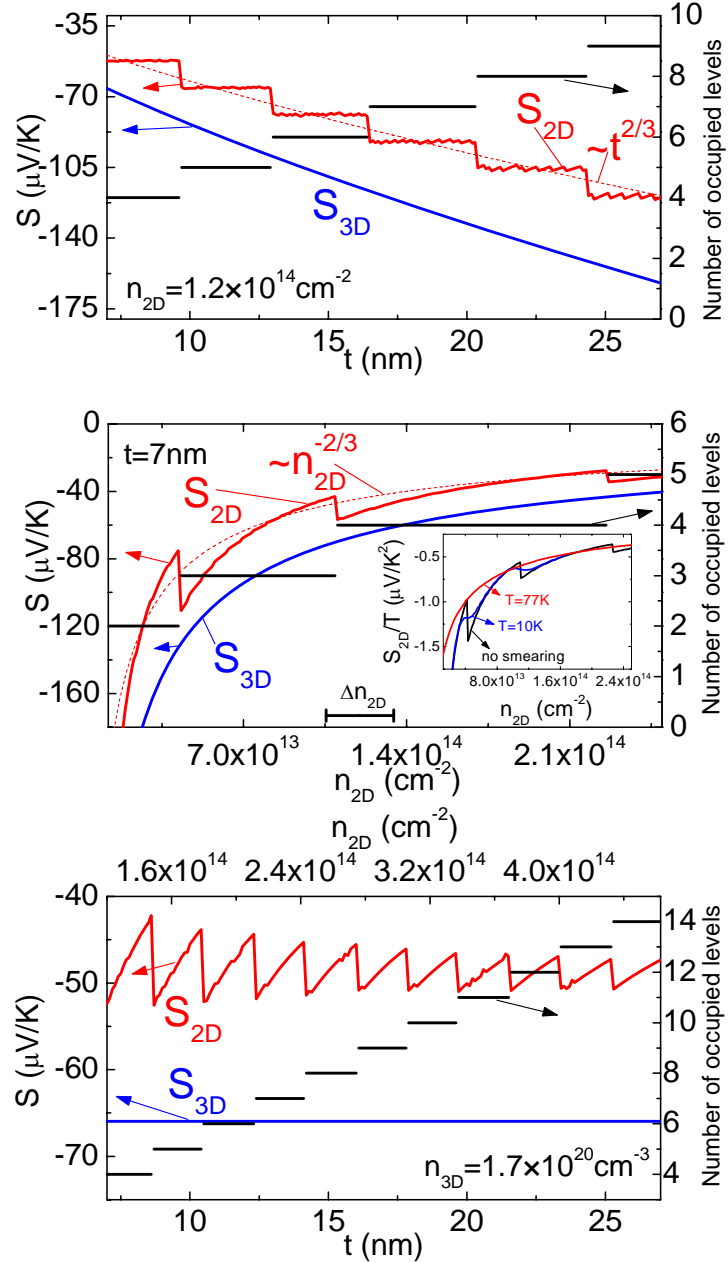

**Figure 1:** (color online) Seebeck coefficients calculated using the 2D approach eq. (8) (red solid lines) and the 3D approach eq. (4) (blue solid lines), for a system characterized by the parameters $m_{eff}=4m_0$ and $\alpha=-0.5$. Power law fitting curves are also shown (red dashed lines). In the right-hand vertical axes, the number of occupied 2D energy levels is indicated. Top panel: dependence on the quantum well width with fixed sheet carrier concentration $n_{2D}=1.2 \cdot 10^{14}$ cm$^{-2}$. Middle panel: dependence on the sheet carrier concentration with fixed quantum well width $t=7nm$; in the main middle panel, the range of $n_{2D}$ that is varied reversibly by field effect in this work is also shown as a bar; inset: same plot as the main panel, calculated using the 2D approach and keeping into account the thermal smearing of the Fermi distribution function at $T=77K$ and $T=10K$. Bottom panel: dependence on the sheet carrier concentration and quantum well width with fixed volume carrier concentration $n_{3D}=1.7 \cdot 10^{20}$ $cm^{-2}$.

Another consideration that emerges form inspection of figure 1 is that confinement, represented by the decrease of the parameter $t$, does not yield any enhancement of the thermopower (see uppermost panel). If confinement is represented by the decrease of the number of occupied levels due to a decrease of $n_{2D}$, a steeper dependence of $S_{2D}$ as a function of $n_{2D}$, compared to $S_{3D}$, is observed within a single quantum step (see middle panel); however, as soon as the next level is reached such gain is compensated by an abrupt step. On the contrary, to obtain thermopower enhancement by 2D confinement starting from a 3D system, a compound with highly anisotropic band structure must be chosen. For example, in the case of anisotropic semiconductors such as $Bi_2Te_3$ and $PbTe$, by suitably choosing the confinement direction with respect to the effective mass tensor, not only the figure of merit Z [21,24] but also the Seebeck coefficient S [25] can be actually improved. Also in systems where the proximity of an interface yields ionic reconstruction and/or band bending, in such a way that the density of states at the Fermi level turns out increased with respect to the bulk, the confinement could be a potential way of enhancing S. Furthermore, lattice strain and deformation may play major roles in affecting the band structure, parametrized by the effective mass, and thus the thermopower; this may be actually the case of experimental systems where confinement is achieved by depositing superlattice structures.

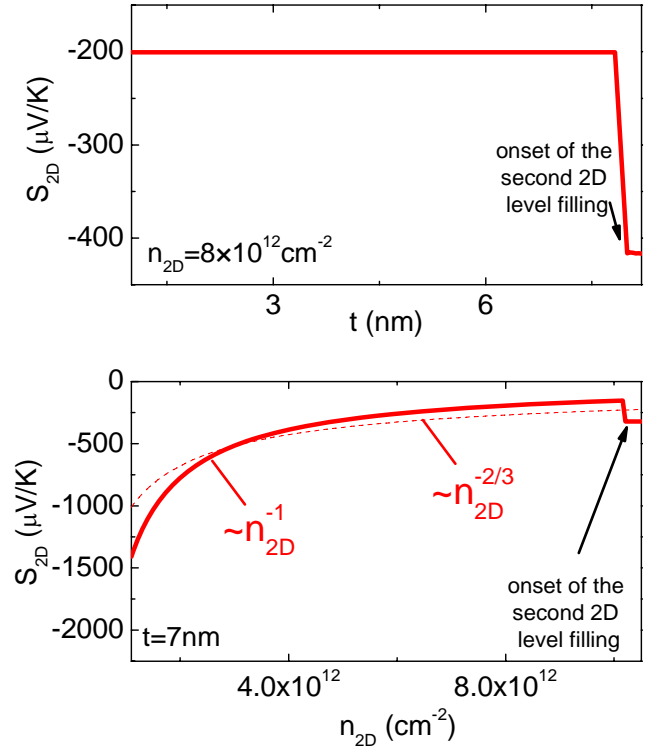

**Figure 2:** (color online) Seebeck coefficients calculated using the 2D approach eq. (8), for a system characterized by the parameters $m_{eff}=4m_0$ and $\alpha=-0.5$; the parameters $t$ and $n_{2D}$ vary within a range where a single 2D energy level is occupied. Top panel: dependence on the quantum well width $t$, with fixed sheet carrier concentration $n_{2D}=8 \cdot 10^{12}$ $cm^{-2}$. Lower panel: dependence on the sheet carrier concentration $n_{2D}$, with fixed quantum well width $t=7nm$; the power law behavior $\propto n_{2D}^{-2/3}$ is also shown for comparison (dashed line).

In figure 1, the calculations are carried out choosing carrier concentration and quantum well width values close to the ones of our experiment, which will be presented and discussed in the following sections. However, it is interesting to see how the thermopower would behave in the extreme 2D limit, that is when only one 2D energy level is occupied. In figure 2, the calculated $S_{2D}$ curves as a function of $t$ for fixed $n_{2D}=8\cdot10^{12}\ cm^{-2}$, and as a function of $n_{2D}$ for fixed $t=7nm$ are shown. Clearly, it can be seen that the power laws $\propto t^{2/3}$ and $\propto n_{3D}^{-2/3}$ do not describe the behaviour of $S_{2D(\nu=1)}$ in the lowest energy level; instead, $S_{2D(\nu=1)}$ depends only on $n_{2D}$, regardless the value of $t$. In particular, $S_{2D(\nu=1)}$ is constant with $t$ in the upper panel of figure 2 and it follows a power law $\propto n_{2D}^{-1}$ in the lower panel. Hence, $S_{2D(\nu=1)}$ is not enhanced by the increasing confinement obtained by decreasing $t$, neither if the quantization steps are averaged out, as noted above, nor in the case where it varies within a given quantum step. On the other hand, it can be said that $S_{2D}$ is increased more steeply by a decrease of the carrier concentration ($\propto n_{2D}^{-1}$) as compared to the 3D case ($\propto n_{2D}^{-2/3}$), if its variability range is limited to a given quantum step.

## 3. Experimental

The samples are prepared by depositing 4 and 6 unit cells of $LaAlO_3$ on $TiO_2$-terminated $SrTiO_3$(001) substrates by PLD, in conditions similar to those of ref. [3]. The substrate temperature and oxygen pressure in the chamber are 770°C and $10^{-4}$ mbar, respectively. After deposition, the samples are annealed for one hour at 550-600°C in 0.2 bar oxygen pressure and then cooled down to room temperature in one hour. A gold pad is evaporated on the back of the 0.5 mm thick substrate and used as a gate electrode, for field effect experiments. For comparison, a Nb doped $SrTiO_3$ (0.5 wt.%) bulk single crystal is measured as well.

Seebeck effect is measured in a home made cryostat, from 77K to room temperature, using an a.c. technique [37]. The sinusoidal period of the power supplied to the sample is 150 s and the applied thermal gradient is around 0.3K across a distance of ~2 mm. Hall effect and resistivity data are measured in a Physical Properties Measurement System (PPMS) by Quantum Design, from 5K to room temperature and in magnetic field up to 9 Tesla.

## 4. Results and discussion

In the upper panel of figure 3, the Seebeck coefficient S measured for the two interfaces and for the bulk Nb doped $SrTiO_3$ sample as a function of temperature is displayed. S is always negative, consistently with electron charge carriers, and an overall linear behavior can be identified in all the curves above 77K. At room temperature, the absolute value |S| is smaller for the two interfaces (400 and 450 µV/K for the 4 unit cell and 6 unit cell $LaAlO_3$ samples ) than in the bulk sample ( 580 µV/K). The linear temperature derivative is also larger for the bulk sample, namely 1.55 µV/K² against ~0.55µV/K² for the two interfaces. Even if the |S| and d|S|/dT values of different samples cannot be directly compared, unless keeping into account the respective carrier concentrations and effective masses, we are brought to the conclusion that no clear enhancement due to confinement is observed in the interfaces. This result is in sharp contrast with some recent experimental findings [29,30], but on the other hand it is fairly plausible, as $SrTiO_3$ is a cubic crystal with modest anisotropy in its electronic properties. In particular, the almost isotropic shape of the Fermi surface has been demonstrated by Shubnikov-de Haas oscillations, which are unchanged for different orientations of the magnetic field [20], as well as by *ab initio* band calculations [47], which indicate that the effective mass along the three (001) directions is $4.4m_0$, while the average effective mass is $4.8m_0$. On the contrary, in $SrTiO_3$-based systems subject to lattice deformation, an enhanced effective mass is indeed expected [47] and consequently the thermopower could turn out to be enhanced [38,26,28] as well.

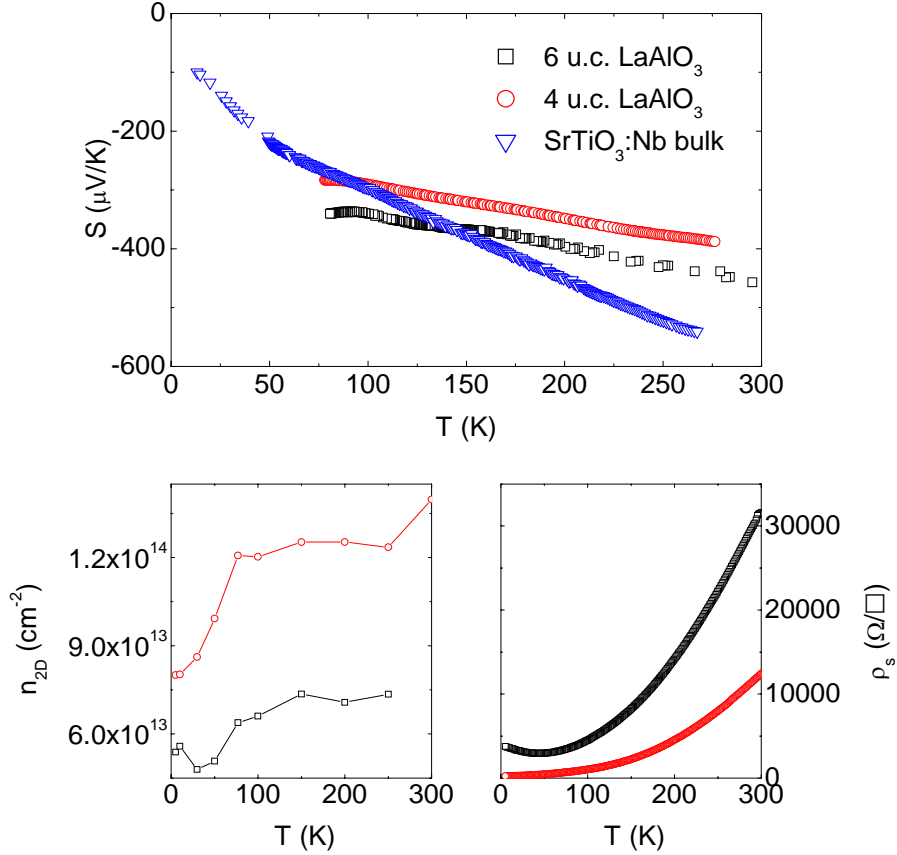

**Figure 3:** (color online) Upper panel: Seebeck coefficient as a function of temperature of two SrTiO$_3$/LaAlO$_3$ interfaces and a bulk Nb doped SrTiO$_3$ crystal. Lower left panel: sheet carrier density of the two SrTiO$_3$/LaAlO$_3$ interfaces extracted by Hall effect. Lower right panel: sheet resistance of the two SrTiO$_3$/LaAlO$_3$ interfaces.

| Sample | dS/dT (μV/K$^2$) | n$_{2D}$ at 77K (cm$^{-2}$) | n$_{2D}$ at 5K (cm$^{-2}$) | ρ$_s$ at 5K (Ω/□) | μ at 5K (cm$^2$V$^{-1}$s$^{-1}$) | ΔS for ΔV$_g$=±200V (μV/K) |
|---|---|---|---|---|---|---|
| LaAlO$_3$/SrTiO$_3$ interface (4 u.c. of LaAlO$_3$) | -0.56 | ~1.2·10$^{14}$ | ~8.0·10$^{13}$ | 240 | 330 | 25 |
| LaAlO$_3$/SrTiO$_3$ interface (6 u.c. of LaAlO$_3$) | -0.55 | ~7.1·10$^{13}$ | ~5.4·10$^{13}$ | 3750 | 31 | 25 |
| Bulk Nb doped SrTiO$_3$ single crystal | -1.55 | ~1.7·10$^{20}$ (cm$^{-3}$) | ~1.8·10$^{20}$ (cm$^{-3}$) | 5.5·10$^{-5}$ (Ωcm) | 630 | - |

**Table I:** parameters of the studied samples: linear temperature derivative of the Seebeck coefficient, carrier concentration at 77K, low temperature sheet resistance/resistivity, low temperature mobility, variation of the measured Seebeck coefficient under field effect.

In the lower panels of figure 3 the other measured transport properties of the two interfaces are presented: on the left-hand side the sheet carrier density extracted from Hall effect data and on the right-hand side the surface resistance. Both samples exhibit metallic behavior, but the 6 unit cell LaAlO$_3$ sample has larger resistance and shows a slight resistance upturn at low temperature, indicative of carrier localization. Consistently, this sample has larger |S| and slightly smaller sheet carrier density, around $7.1 \cdot 10^{13}$ cm$^{-2}$ at high temperature, against the value around $1.2 \cdot 10^{14}$ cm$^{-2}$ of the 4 unit cell LaAlO$_3$ interface. Regarding the different transport properties of the two samples, it must be said that the carrier density of the electron gas present at the LaAlO$_3$/SrTiO$_3$ interface may vary up to a factor five for the same LaAlO$_3$ thickness and deposition parameters. The different $n_{2D}$ of the two samples reported in this paper are thus within the range of variability observed in different samples prepared in these conditions. Indeed, a recent report on the dependence of the transport properties for the LaAlO$_3$/SrTiO$_3$ system shows that below 10 unit cells of LaAlO$_3$, $n_{2D}$ is not determined by the thickness of the LaAlO$_3$ layer [39]. The measured parameters of each sample are summarized in table I. Due to the overall similarity of the properties of the two interfaces, in the following, we will carry out quantitative analysis on one only, namely the 4 unit cell LaAlO$_3$ sample.

For SrTiO$_3$, it has been evidenced that the phonon drag contribution to thermoelectric power above the quantum paraelectric Curie temperature ~40K is negligible [40,41], thereby the model of diffusive thermopower illustrated in section 2 should apply. However, in order to analyze quantitatively the experimental data, we point out that all the above equations describe the linear temperature dependence of the diffusive contribution to S, provided that the carrier density remains constant with temperature and no change in the scattering mechanisms occurs. This condition certainly applies to our samples in the temperature range between 77K and 300K. At lower temperature different electrical and transport mechanisms occur, for example phonon drag, temperature dependence of the carrier density, electron scattering mechanisms other than by acoustic phonons. Thereby, at lower temperature S cannot be described by a linear temperature law anymore; instead, it eventually vanishes with decreasing temperature with a steeper than linear dependence [30,40,41]. The low temperature behavior affects the absolute values of the diffusive S in the temperature window 77K-300K by an "offset", so that the above equations must be used to fit not just S but its changes with temperature and carrier density ($\partial S/\partial T$ and $\partial S/\partial n_{2D}$), which are free from any unknown "offset".

We try to apply both the 3D and 2D analyses in order to find out which of them is the most appropriate. We fix the parameter $\alpha \approx -0.5$, for scattering by acoustic phonons [40,34,42]. For the Nb doped bulk sample, using carrier density data from Hall effect, which is $n_{3D} \approx 1.7 \cdot 10^{20}$ cm$^{-3}$ at high temperature, we extract directly the value of the effective mass $m_{eff} \approx 7.2 m_0$. This result in perfect agreement with data measured in similar samples [43]. For the interfaces the outcome is not univocal because the thickness $t$ of the conducting interface is a further unknown parameter. Literature values of the SrTiO$_3$ effective mass range between $1.1 m_0$ and $13 m_0$ [40,33,44,45,46,41]. Relying on the assumption that for SrTiO$_3$ $m_{eff}$ cannot be realistically smaller that the free electron value, we extract upper limit values for the thickness, corresponding to the lower limit value of the effective mass $m_{eff} \approx 1.1 m_0$. Using the 3D approach, we obtain that the conducting interface is thinner than $t_{max} \approx 25$ nm, while using the 2D approach we get $t_{max} \approx 40$ nm; the latter value corresponds to *12* occupied energy levels. A lower limit for $t$, $t_{min} \approx 1$ nm, comes out by assuming $m_{eff} \approx 13.5 m_0$ in the 2D approach, which corresponds to one single occupied level. On the other hand, a more plausible estimation of the thickness $t$ is found with the $m_{eff}$ experimentally found in SrTiO$_3$ samples $m_{eff} \approx 4 m_0$ [33,46,47], which yields *t≈3.8 nm* and *t≈6.8 nm* in the 3D and 2D approaches, respectively. These small values of $t$ clearly indicate that a 2D approach is more suitable for our systems; indeed *t≈6.8 nm* corresponds to only *4* occupied levels. This finding about the 2D character of LaAlO$_3$/SrTiO$_3$ interfaces is in agreement with the 2D character of superconductivity found in similar samples [3]. In fact, in our case the requirement is even more stringent as long as superconductivity of 2D character is observed in systems whose thickness is smaller than the coherence length $\xi \approx 70$-$100$ nm, while in our case the signature of a 2D-like density of states implies that the system has a small number of occupied levels, which in general occurs at even smaller thicknesses. In the following, only the 2D equations will be used; however the 3D approach gives similar results. The difference in effective masses of Nb-doped bulk sample and of the interfaces is not surprising, as Nb substitution is theoretically predicted [47] and experimentally observed [43] to enhance the effective mass. Moreover, in the interfaces, the average effective mass results from the curvature of the cross-section of the slightly anisotropic Fermi surface [47] and could turn out to be smaller than the average bulk value. To give an idea of the variability range of $t$ and $m_{eff}$, we mention that for $t \approx 10nm$ ($t \approx 5nm$), taken from ref. [3,16,18], we obtain $m_{eff} \approx 2.7 m_0$ ($m_{eff} \approx 4.5 m_0$), corresponding to 5 (3) occupied levels.

We now turn to the dependence of S on the carrier density. This effect could be explored by analyzing a large series of samples; however, different interface samples inevitably have different degrees of disorder, which makes a fine comparison unreliable. Moreover, a restricted range of deposition parameters must be used to fabricate these interfaces, to avoid the parallel transport in the deoxygenated SrTiO$_3$ substrate [48], thus making it almost impossible to vary safely only the carrier concentration. We circumvent this hindrance by measuring directly the dependence of S on the back gate voltage, thus tuning reversibly the carrier concentration by field effect. Also surface resistance and Hall effect are measured under field effect, so that the dependence of S on the measured carrier concentration is directly obtained. Moreover, since surface resistance and Hall resistance are related to the sheet carrier density $n_{2D}$, while S is related to the volume carrier density $n_{3D}$, this approach allows to extract information of the conducting thickness $t$, as well as on the spatial distribution of the charge carrier, upon application of a back gate voltage.

In figure 4 we present the measured Seebeck coefficient for the two interfaces as a function of the carrier concentration, tuned by field effect with a gate voltage of ±200V, corresponding to an electric field of ±4MV/m. In the bottom horizontal axis is the carrier concentration measured by Hall effect under field effect, while in the top horizontal axis is the applied gate voltage; the relation between these two quantities is linear, as displayed by the Hall effect data measured in the 4 LaAlO$_3$ unit cell interface, in the bottom left panel of figure 4. In the bottom right panel, the corresponding change of the sheet resistance $\rho_s$ by field effect for the 4 LaAlO$_3$ unit cell interface is shown. As expected, a positive gate voltage accumulates negative charge carriers, thus increasing $n_{2D}$ and decreasing $\rho_s$, while a negative gate voltage causes carrier depletion and increases $\rho_s$. Surprisingly, in figure 4 we observe that the absolute value of the Seebeck coefficient increases with increasing back gate voltage, which is seemingly in contrast with what expected from the inverse dependence of |S| on the carrier concentration (see middle panel of figure 1). To account for this result, a simple explanation on terms of leakage current from the gate electrode is ruled out. Indeed the leakage current is monitored during each measurement; it is negligibly small at 77K and becomes few hundreds of pA at room temperature. Only in the latter case it results in a detectable contribution to the Seebeck voltage $S(V_g) \approx S(-V_g)$.

In order to sort out this puzzle, we have to consider the spatial distribution of charge carriers. The back gate voltage rises and lowers the potential barrier that confines the conducting interface layers, thus changing not only the total number of carriers, but also the width of the potential well. This situation is schematically sketched in figure 5, where a qualitative picture of the conduction band bending and charge density profile for different values of back gate voltages is shown. If the widening of the potential well due to a positive gate voltage is large enough, it happens that the average volume carrier density in the potential well $\overline{n_{3D}} = t^{-1} \int n_{3D}(z)dz$ decreases, even if its integral along the thickness (spatial direction $z$) $n_{2D} = \int n_{3D}(z)dz$ increases. If $\overline{n_{3D}}$ decreases, |S| increases even in a 2D system, as $S_{2D}$ has the same average functional dependence on the volume carrier density as $S_{3D}$ in eq. (4) (see figure 1). The integral $n_{2D} = \int n_{3D}(z)dz$ is indeed measured by Hall effect under applied gate voltage, thereby we can estimate the widening of the confining well by the back gate voltage, combining Hall effect and Seebeck effect data. By differentiation we can write:

$$\Delta n_{2D} = \Delta(\overline{n_{3D}}t) \approx t \cdot \Delta \overline{n_{3D}} + \overline{n_{3D}} \Delta t \approx t \cdot \Delta \overline{n_{3D}} + \frac{n_{2D}}{t} \Delta t \qquad (9)$$

Here, the quantities $\Delta n_{2D}$ and $n_{2D}$ are measured by Hall effect with and without field effect, while $\overline{\Delta n_{3D}}$ is extracted from the variation of the Seebeck effect under field effect. We note that for a positive variation of back gate voltage, in the right-hand side of eq. (9), the term $t \cdot \overline{\Delta n_{3D}}$ is negative (mechanism of charge dilution), while the term $(\Delta n_{2D}/t)\cdot \Delta t$ is positive (mechanism of quantum well widening). A proper (not univocal) choice of $t$ and $\Delta t$ fulfills eq. (9). Assuming $m_{eff}=4m_0$ and $t\approx 6.8nm$ as above, we obtain $\Delta t\approx 6.2$ nm. In other words, the system passes from 4 to 6 occupied levels by field effect. In figure 4, the dashed line is calculated as the variation $|\Delta S|\approx -6.4\mu V/K$ expected from the variation of $\Delta n_{2D}$ measured by Hall effect under field effect for the 4 LaAlO$_3$ unit cell sample, assuming a constant quantum well width. Clearly its behavior is opposite to the measured one. Instead, the continuous line is calculated from eq. (8) and (9), keeping into account both the changes $\Delta n_{2D}$ and $\Delta t$, which gives the correct trend $|\Delta S|\approx +25\mu V/K$.

As seen in the qualitative sketch of figure 5, the charge dilution mechanism occurs because the wall of the potential well on the side of the SrTiO$_3$ substrate is shallow; this is actually a pretty realistic picture, as the Fermi level in the bulk SrTiO$_3$ shifts away from the middle gap toward the conduction band edge whenever a tiny amount of oxygen vacancies is present [49]. In PLD deposited samples, oxygen vacancies are very likely to form, even if they may be present in negligible amounts to be observed by transport measurements. Moreover, the reliability of the qualitative picture sketched in figure 5 is confirmed by the charge profile recently measured by infrared ellipsometry [50], which is shown to have a strongly asymmetric shape with a rapid initial decay over the first 2 nm and a pronounced tail that extends to about 11 nm. Indeed, such depth profile of charge density, just expected when the wall of the potential well on the side of the SrTiO$_3$ substrate is shallow, is particularly liable to the above described charge dilution mechanism by a back gate voltage.

Finally, it is worth noticing that a close inspection of experimental data of figure 4 evidences that for positive gate voltages the slope of the S curve tends to saturate in both interfaces; this may be an indication that the charge dilution mechanism eventually saturates and a further increase of gate voltage results in an increase of volume carrier density $\overline{n_{3D}}$, with virtually constant width of the conducting layer.

We note that no quantization effects similar to those in the middle panel of figure 1 are seen in figure 4. One reason may be that the range of $n_{2D}$ spanned by field effect is too small, so that the number of occupied levels remains unchanged and S varies continuously. In the middle panel of figure 1, the bar corresponding to the range $\Delta n_{2D}$ is indicated. Another reason may be smearing by finite temperature. In the inset of figure 1, the Seebeck coefficient curves calculated by keeping into account thermal smearing suggest that this effect should play a major role at 77K. Indeed, in these systems, even if the thermal energy at 77K is ~3 times smaller than the spacing between the levels adjacent to the Fermi level, it is not evident whether the measuring temperature is smaller than the Fermi temperature $T_F$, which is required for quantization features to be resolved. $T_F$ may vary in a range as large as ~2K to ~100K with increasing band filling of a single level (this is actually the range spanned by $T_F$ for ν=4 occupied levels, $m_{eff}=4m_0$ and $t$ varying form *6.8nm* to *9.5nm)*; hence it cannot be estimated *a priori* whether $T_F$ is indeed larger that 77K. In any case, no quantization steps are visible in conductance measurements under field effect down to 5K, either.

We also suggest that alternatively to the above model of widening of the potential well by $\Delta t$, the effect of the back gate voltage could be of tuning charge density in a portion of the bulk SrTiO$_3$ substrate, which would contribute to transport properties in parallel with the interface charge. In the case of $\rho_s$, the bulk parallel contribution might be negligible, due to its much smaller conductivity $\sigma_{bulk}<<\sigma_{interface}$ (a value $\sigma_{bulk}\approx 10^{-5}\sigma_{interface}$ has indeed been measured [18]), but it may be detectable in the measured thermopower:

$$S = \frac{\sigma_{interface} S_{interface} + \sigma_{bulk} S_{bulk}}{\sigma_{interface} + \sigma_{bulk}} \qquad (10)$$

If $S_{bulk} > S_{interface}$, the measured S could actually increase in absolute value with increasing gate voltage, as observed.

As a final comment to our results, it must be said that the present simplified approach that assumes an effective width within which the charge carrier density is uniform may be inadequate to describe the system quantitatively, therefore the precise numerical results on the widening of the conducting layer should be taken with caution. Indeed, it is likely that the charge profile is strongly non uniform [50], so that measured S, $\rho_s$ and Hall resistance $R_{Hall}$ are the results of integrals along the sample depth $z$ of the charge profile, which cannot be easily extracted.

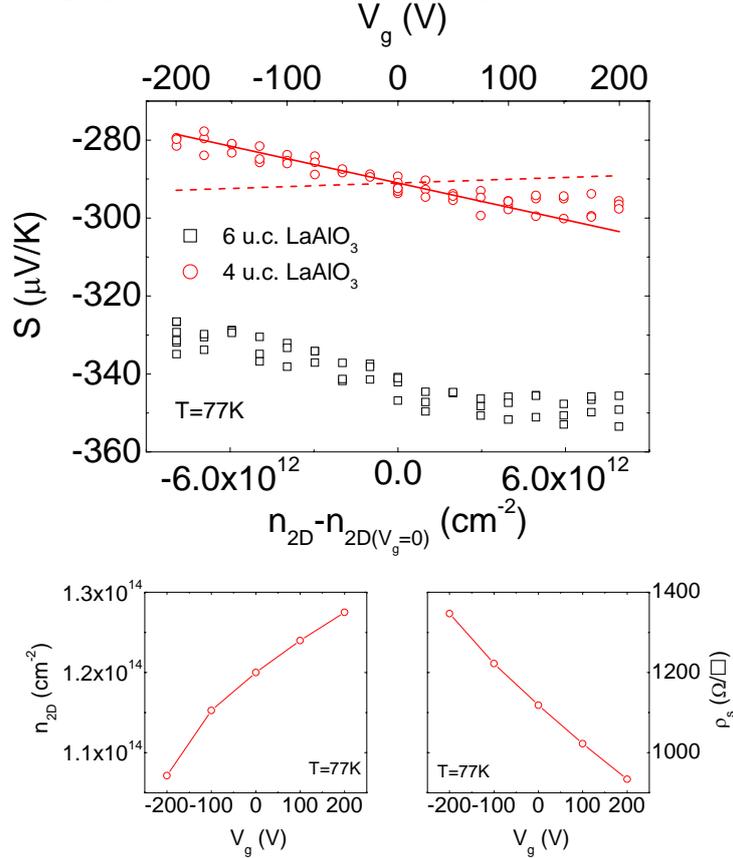

**Figure 4:** (color online) Upper panel: Seebeck coefficient as a function of measured carrier density and back gate voltage for the two $SrTiO_3/LaAlO_3$ interfaces at 77K. Lower left panel: sheet carrier density of the 4 unit cell $LaAlO_3$ interface extracted by Hall effect under field effect. Lower right panel: sheet resistance of the 4 unit cell $LaAlO_3$ interface under field effect.

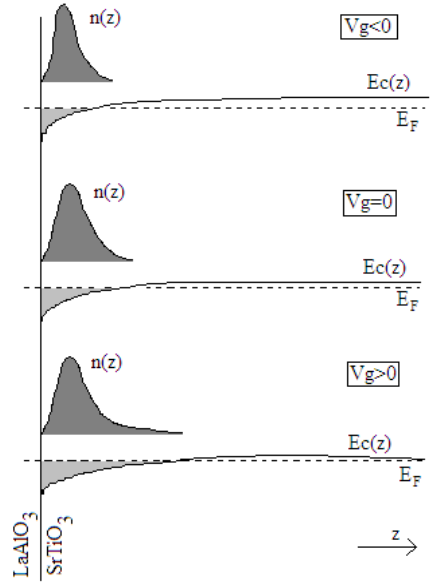

**Figure 5:** Qualitative sketch of the conduction band bending and charge density profile for different back gate voltages.

## 5. Conclusions

We carry out a complete electrical and thermoelectrical characterization of $LaAlO_3/SrTiO_3$ interfaces; in particular, we measure conductivity, Hall effect and Seebeck effect as a function of temperature and carrier concentration, tuned by field effect. By combining these data, we get information on the spatial extension of the conducting layer, as well as on the effect of a back gate voltage on the charge distribution. Indeed, the Seebeck coefficient is related to the volume carrier concentration even for 2D systems, whereas the other measured transport properties give information on the sheet carrier density.

A quantitative analysis of thermopower experimental data using either 3D or 2D frameworks indicates that the system is described by a density of states of 2D character and it is confined in a *7nm* wide well, with *4* occupied levels at 77K.

No enhancement of the thermopower by quantum confinement with respect to bulk sample is visible, consistently with the almost isotropic band structure of $SrTiO_3$ and the absence of significant lattice deformation.

A back gate voltage allows to tune both carrier concentration and its spatial distribution; in particular, we find that the width of the potential well where the mobile carrier are confined is widened from 7 to 13 nm by a gate voltage of ±200V. As for the sheet carrier density , it is varied in a range $\Delta n_{2D} \sim 1.5 \cdot 10^{13}$ $cm^{-2}$ by the same back gate voltage, which is too limited to observe quantization steps in the thermopower plot. Besides, thermal smearing at 77K certainly plays a major role in blurring out any such steps. Hence, the back gate seems not to be a suitable tool to investigate quantum effects. Given the limited and critical range of deposition parameters necessary to prepare these samples [9,48], the possibility of fabricating interfaces with a carrier density low enough to observe quantum effects appears to be a difficult task, too. Alternatively, top side-gates and planar patterning could be promising tools to achieve possibly the quantum limit of one single 2D level occupied in $LaAlO_3/SrTiO_3$ interfaces.